# Introducing First-Principles Calculations: New Approach to Group Dynamics and Bridging Social Phenomena in TeNP-Chain Based Social Dynamics Simulations

Yasuko Kawahata [†]

Faculty of Sociology, Department of Media Sociology, Rikkyo University, 3-34-1 Nishi-Ikebukuro,Toshima-ku, Tokyo, 171-8501, JAPAN.

ykawahata@rikkyo.ac.jp

**Abstract:** This note considers an innovative interdisciplinary methodology that bridges the gap between the fundamental principles of quantum mechanics applied to the study of materials such as tellurium nanoparticles (TeNPs) and graphene and the complex dynamics of social systems. The basis for this approach lies in the metaphorical parallels drawn between the structural features of TeNPs and graphene and the behavioral patterns of social groups in the face of misinformation. TeNPs exhibit unique properties such as the strengthening of covalent bonds within telluric chains and the disruption of secondary structure leading to the separation of these chains. This is analogous to increased cohesion within social groups and disruption of information flow between different subgroups, respectively. . Similarly, the outstanding properties of graphene, such as high electrical conductivity, strength, and flexibility, provide additional aspects for understanding the resilience and adaptability of social structures in response to external stimuli such as fake news. This research note proposes a novel metaphorical framework for analyzing the spread of fake news within social groups, analogous to the structural features of telluric nanoparticles (TeNPs). We investigate how the strengthening of covalent bonds within TeNPs reflects the strengthening of social cohesion in groups that share common beliefs and values. The breakdown of the secondary structure of TeNP can be likened to the fragmentation of information flows between subgroups with different beliefs, resulting in information silos. Introducing the concepts of resonance and amplification, we model how fake news resonates with specific subgroups, leading to diffusion and amplification within those circles. The separation of telluric chains within nanoparticles will be used as an analogy for how certain subgroups adhere adamantly to misinformation and thus isolate themselves from the broader social discourse.

## 1. Introduction

article In this research note, we discuss the application of first-principles calculations methods to social simulations, particularly focusing on the structural characteristics of tellurium nanoparticles and their interaction with the spread of fake news, as well as whether tellurium chains become isolated chains.

Research has shown that (1) tellurium chains formed by covalent bonds persist while the covalent bonds become stronger, and (2) the interactions between chains in the secondary structure collapse with the formation of tellurium nanoparticles, leading to the breakdown of interactions between basic structures and the isolation of tellurium chains. We aim to validate these findings using first-principles calcu-lations and apply the same methodology to represent social groups or subgroups, especially sensitive to fake news. We will explore how these groups interact with fake news and amplify its impact. Finally, we will discuss methods for analyzing whether tellurium chains become isolated chains. We will then discuss the potential application of first-principles calculation methods in the field of condensed matter physics to social simulations.

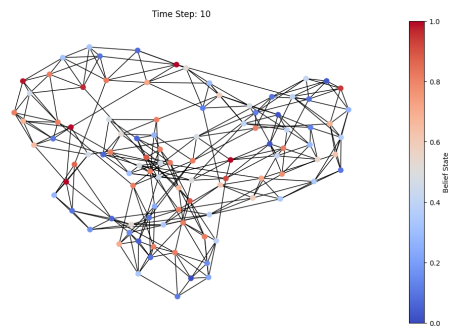

Fig. 1: Network Belief States susceptibility

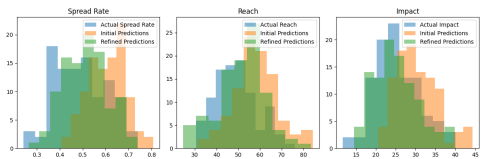

Fig. 2: Actual Reach, Speed, Spread

In the contemporary digital landscape, the rapid proliferation of fake news poses significant challenges to societal cohesion and informed decision-making. Traditional models of understanding misinformation spread often focus on psychological and sociological factors, overlooking the potential insights that could be gained from interdisciplinary approaches. This paper introduces a novel metaphorical framework that bridges nanoscience and social science, specifically leveraging the structural characteristics of tellurium nanoparticles (TeNPs) to analyze the dynamics of fake news spread within social groups.

Tellurium, a metalloid element, forms intricate nanostructures characterized by strong covalent bonds within tellurium chains and weaker interchain interactions. This dual nature of bonding within TeNPs serves as an apt metaphor for the intricate dynamics of social cohesion and fragmentation in the face of misinformation. Just as the strength of covalent bonds within tellurium chains influences the stability and properties of the nanoparticles, the strength of shared beliefs and values within social groups plays a crucial role in their susceptibility to misinformation. Similarly, the breakdown of secondary structures in TeNPs, leading to isolated tellurium chains, mirrors the disruption of information flow between subgroups with divergent beliefs, fostering informational silos and echo chambers.

By drawing parallels between the atomic-level interactions within TeNPs and the social interactions that govern the spread of fake news, this paper seeks to shed light on the underlying mechanisms that facilitate the resonance and amplification of misinformation within certain subgroups. Through this interdisciplinary lens, we aim to not only enhance our understanding of the social dynamics at play in the era of fake news but also to explore novel strategies for mitigating its impact on society. The following sections will delve into the theoretical underpinnings of our framework, outline our methodology for simulating the spread of fake news using this model, and discuss the implications of our findings for both nanoscience and the broader social science community in addressing the challenges posed by misinformation.

In recent years, the rapid proliferation of misinformation, particularly in the form of fake news, has emerged as a formidable challenge with far-reaching implications for society. The complexity and dynamism of social interactions, compounded by the intricate nature of human beliefs and behaviors, render traditional analytical approaches insufficient for comprehensively understanding and addressing the spread of misinformation. This paper introduces an innovative interdisciplinary methodology that bridges the gap between the fundamental principles of quantum mechanics, as applied in the study of materials like tellurium nanoparticles (TeNPs) and graphene, and the intricate dynamics of social systems.

The foundation of this approach lies in the metaphorical parallel drawn between the structural characteristics of TeNPs and graphene and the behavioral patterns of social groups in the face of misinformation. TeNPs exhibit unique properties, such as the strengthening of covalent bonds within tellurium chains and the breakdown of secondary structures leading to the isolation of these chains, which we analogize to the increased cohesion within social groups and the disruption of information flow between distinct subgroups, respectively. Similarly, the exceptional properties of graphene, including its high electrical conductivity, strength, and flexibility, offer additional dimensions for understanding the resilience and adaptability of social structures in response to external stimuli like fake news.

By leveraging first-principles calculations, a methodology rooted in the fundamental laws of quantum mechanics, this paper explores the potential of these computational techniques to simulate and predict social dynamics, particularly the dissemination patterns of misinformation. The aim is to transcend the conventional boundaries of social sciences by incorporating insights from physical sciences, thereby providing a more nuanced and predictive understanding of social phenomena.

This interdisciplinary fusion not only enriches our comprehension of social dynamics through the lens of quantum mechanics but also opens new avenues for developing targeted interventions to mitigate the spread of misinformation. Through a detailed exploration of the metaphorical relationships between the properties of TeNPs, graphene, and social group behaviors, this paper lays the groundwork for a novel theoretical framework that holds promise for advancing the field of social simulations and, ultimately, fostering more resilient and informed societies.

# 2. Potential Applications of First Principles-based Physical Science Methods in Social Simulation

Applying first principles calculations from physics to social simulations opens up new possibilities in the field of social science, but it also comes with several considerations. Below, we propose hypotheses regarding social simulations inspired by the study of graphene's properties, as well as discuss the potential and considerations associated with their application.

## 2.1 Potential Applications to Social Simulation

- **Understanding Complex Systems:** First principles calculations have the power to elucidate complex physical phenomena from fundamental principles. Applying this to social simulation may reveal the basic mechanisms of complex social systems such as individual and group behaviors and social interactions.

- **Construction of Precise Predictive Models:** First principles calculations in physics can lead to highly accurate predictions in certain cases. Applying this method to model social phenomena can enhance the predictive accuracy of social phenomena.

- **Enhancement of Quantitative Analysis:** Social simulations based on first principles calculations can strengthen quantitative research methods in social science. Applications in fields like sociophysics and computational social science, dealing with large datasets, are particularly conceivable.

## 2.2 Considerations

- **Applicability of Analogies:** There are fundamental differences between physical and social phenomena. Applying laws and parameters from physical phenomena to social phenomena is challenging, requiring careful interpretation.

- **Validity and Limits of Models:** Modeling social phenomena from first principles is extremely difficult. Social factors are diverse, and many elements, such as individual consciousness and cultural backgrounds, cannot be explained solely by physical laws.

- **Ethical and Social Impacts:** Precise social simulations may raise ethical issues regarding individual privacy and social interventions. Careful consideration is required regarding how to handle the results of simulations.

## 2.3 Hypotheses for Social Simulation Inspired by Graphene's Properties

The unique properties of graphene, such as its robust structure and excellent electrical conductivity, can be applied as analogies to social phenomena, such as efficiency of information transmission and stability of social structures. For instance, modeling social networks based on the hexagonal lattice structure of graphene and modeling the strength of connections and efficiency of information flow may provide insights into optimal social system structures and information transmission optimization.

One hypothesis could be that "social networks similar to the lattice structure of graphene exhibit superior efficiency and resilience in information transmission." Through such models, principles for designing efficient and stable social systems can be explored.

# 3. Indirect Applications of First Principles Calculations to Social Simulation

Research directly applying first principles calculations to social simulation is still rare, and there are limited examples of such prior studies. However, there are several instances where theories and methods from physics and chemistry have been indirectly applied to social science and economics. Below, we enumerate some of these indirect applications:

- **Application of Spin Glass Theory to Economics**: Spin glass theory, a theory in physics concerning the magnetism of materials, has been applied to economic systems characterized by uncertainty and competing interactions. Research utilizing this theory analyzes dynamics in financial markets and optimization problems in decision-making.

- **Application of Ising Model to Social Phenomena**: The Ising model, a physics model describing the magnetization of materials, has been applied to sociology and economics as a model for collective decision-making and opinion formation. It is used, for example, to analyze how individual opinions change through social interactions.

- **Application of Chaos Theory to Traffic Flow**: Chaos theory, a theory studying nonlinear dynamic systems where small differences in initial conditions lead to significant differences in outcomes, has been applied to predict and optimize traffic flow. It contributes to understanding the mechanisms of traffic congestion and designing efficient transportation systems.

- **Application of Network Theory to Infectious Disease Models**: Network theory in physics models various systems as networks and analyzes their structure and dynamics. This theory is applied to model how infectious diseases spread through contact networks of people and information, aiding in predicting and devising strategies for preventing infectious diseases.

These examples demonstrate how theories from physics and chemistry can contribute to understanding and predicting social phenomena. However, these applications are not directly derived from "first principles calculations" but rather adapt theories and models from physics and chemistry to address problems in social science. Direct applications of first principles calculations to social simulation await further development in future research.

The application of first-principles calculations, traditionally used in material science to understand the properties of substances like graphene, to the domain of social simulations, particularly in the context of fake news dissemination, offers a promising and innovative approach. Graphene, known for its exceptional electrical conductivity, mechanical strength, and flexibility, provides an intriguing analogy for exploring the dynamics of information flow, resilience, and adaptability within social networks.

By leveraging the principles underlying graphene's unique properties, we hypothesize that social simulations can be significantly enhanced to more accurately predict and understand the patterns and impacts of fake news dissemination across social networks. Specifically, we propose the following potential insights and applications:

Electrical Conductivity and Information Flow, Graphene's high electrical conductivity can be metaphorically related to the efficient and rapid spread of information within social networks. Just as electrons move freely and swiftly across graphene's two-dimensional lattice, information, including fake news, can traverse social networks with ease, facilitated by interconnected nodes (individuals) and edges (social connections). Simulating social networks with the analogy of graphene's conductivity could offer insights into how quickly and widely fake news can spread, identifying potential 'high-conductivity' pathways or nodes that contribute to faster dissemination.

Mechanical Strength and Resilience, The extraordinary strength of graphene, attributed to its hexagonal lattice structure, can be paralleled with the resilience of certain social groups or beliefs systems against misinformation. In simulations, this could be represented by the robustness of certain network clusters or the presence of 'strong bonds' within certain communities that resist the penetration or spread of fake news. Understanding these resilient structures could help in designing strategies to strengthen societal resilience against misinformation.

Flexibility and Adaptability, Graphene's flexibility, despite its strength, offers a metaphor for the adaptability of social networks to changing information landscapes. In the face of new or conflicting information, social groups, like graphene, may exhibit both rigidity (resistance to fake news) and flexibility (adaptability to new, accurate information). Simulating this dual nature could provide insights into how misinformation campaigns adapt over time and how social networks can evolve to counteract their effects.

Edge Effects and Peripheral Influences, In graphene, properties such as edge states can significantly affect its overall behavior. Similarly, in social networks, individuals or groups at the periphery can have disproportionate effects on the spread of misinformation. Simulating these 'edge effects' can shed light on how peripheral nodes that may seem inconsequential can serve as critical points for the introduction or amplification of fake news.

The hypothesis posits that by applying first-principles calculations and the metaphorical insights derived from the properties of graphene, social simulations of fake news dissemination can achieve greater predictive accuracy and depth of understanding. This interdisciplinary approach not only provides a novel lens to view the complexities of social networks but also paves the way for developing more effective interventions to combat the spread of misinformation in our increasingly connected world.

# 4. Application of Tellurium Nanoparticles and Graphene to Plasmonics and Hypotheses for Social Simulation

Tellurium nanoparticles and graphene have attracted attention in the field of plasmonics due to their unique physical properties. Plasmonics refers to the phenomenon where free electrons in metals collectively oscillate in response to external stimuli such as light. This collective oscillation leads to the enhancement of local electromagnetic fields, holding potential applications in sensors and optical communication.

## 4.1 Tellurium Nanoparticles and Plasmonics

Tellurium is a semiconductor material, and its properties at the nanoscale are under research. Tellurium nanoparticles may induce localized surface plasmon resonance (LSPR) due to their unique structures. LSPR, regulated by the size, shape of nanoparticles, and surrounding dielectric constant, makes the properties of tellurium nanoparticles crucial in modulating these plasmonic characteristics. This property can be utilized in various applications such as highly sensitive sensors and novel types of solar cells.

## 4.2 Graphene and Plasmonics

Graphene is a two-dimensional material consisting of carbon atoms with unique electronic properties. Plasmons in graphene have garnered particular attention due to its high carrier mobility and tunable electronic properties. Plasmons in graphene occur over a wide frequency range from infrared to visible light, and their resonance frequency can be tuned by the electric field effect. This tunability enables the development of highly customizable optical devices and sensors.

### 4.3 Hypotheses for Application in Social Simulation

The idea of applying the plasmonic properties of tellurium nanoparticles and graphene to social simulation is based on likening the transmission and diffusion of information to the electronic properties of materials. For instance, modeling the transmission of information and resonance within social networks as phenomena similar to plasmon resonance is conceivable.

- **Enhancement of Information Transmission**: Just as plasmon resonance enhances local electromagnetic fields, specific nodes or groups within social networks may function as amplifiers of information. Identifying such nodes or groups and understanding their characteristics can facilitate the design of effective transmission and diffusion strategies for information.

- **Tunable Information Transmission**: Similar to the tunability of plasmon resonance in graphene through external stimuli, the mechanisms of information transmission in social networks may also be adjustable through external stimuli or interventions. Understanding conditions where specific messages or policies induce social resonance can enable strategies to maximize social impact.

- **Understanding Nonlinear Dynamics**: Just as plasmonic properties exhibit nonlinear optical responses, social interactions may also demonstrate nonlinear characteristics. Understanding this nonlinearity can predict unexpected large-scale social changes and diffusion patterns of information, allowing for appropriate measures to be taken.

In conclusion, by analogizing the physical properties of tellurium nanoparticles and graphene to social simulation applications, new theoretical frameworks for information transmission can be provided, leading to a deeper understanding and control of social processes.

First-principles calculations that explore the physical properties of advanced materials such as tellurium nanoparticles (TeNPs) and graphene are powerful methods for detailed analysis of electron behavior and the interaction of light within materials. These calculations play a vital role in new discoveries and technological advances in materials science. Interestingly, the application of such computational methods for physical phenomena to problems in the social sciences, particularly in understanding the mechanisms of the spread of fake news, opens the way to understanding the dynamics of social systems from a new perspective.

The proliferation of fake news in modern society has a significant impact on the formation of individual and group opinions, sometimes even causing social divisions. By comparing the process of fake news diffusion to phenomena such as TeNPs or plasmon resonance in graphene, it is possible to mathematically model how information resonates and is amplified among individual people and groups. This approach allows for theoretical analysis of the conditions under which fake news tends to resonate strongly within certain groups, the refraction and reflection of information between groups with different opinions, and the formation of filter bubbles.

In addition, the model provides a framework for identifying the interactions among the spreaders of fake news and the process of social isolation that occurs due to information bias. It is possible to quantify how disseminators receive and share information, how they reinforce or change their beliefs in the process, and how this affects the overall social information environment. These insights are expected to contribute to the development of specific strategies to curb fake news and filter bubbles.

By applying computational methods in condensed matter physics to the analysis of social systems, we propose a new theoretical framework for understanding the diffusion of fake news and its social consequences. By clarifying the mechanisms of interactions between groups and isolation associated with the proliferation of fake news, we aim to take a step toward the creation of a healthy information environment.

Utilizing methods from physics based on first-principles calculations in social simulations opens up new possibilities in the field of social sciences, but it also involves several considerations that need attention. Below, we propose the potential, considerations, and hypotheses for social simulations inspired by the study of graphene's properties.

## 5. Potential for Application in Social Simulations

- **Understanding Complex Systems**:

    First-principles calculations have the power to elucidate complex physical phenomena from fundamental principles. Applying this to social simulations could potentially reveal the basic mechanisms of complex social systems such as individual and group behaviors and social interactions.

- **Building Precise Predictive Models**:

    First-principles calculations in physics sometimes allow for highly accurate predictions. By applying this method to model social phenomena, it may be possible to improve the predictive accuracy of social phenomena.

- **Enhancing Quantitative Analysis**:

Social simulations based on first-principles calculations can strengthen quantitative research methods in social sciences. This could particularly be relevant in fields like sociophysics and computational social science dealing with large datasets.

0 **Applicability of Analogies**:

There are fundamental differences between physical and social phenomena. Applying laws and parameters of physical phenomena directly to social phenomena is challenging and requires careful interpretation.

0 **Validity and Limitations of Models**:

Modeling social phenomena from first principles is extremely difficult. Social factors are diverse, and many elements such as individual consciousness and cultural backgrounds cannot be explained solely by physical laws.

0 **Ethical and Social Implications**:

Precise social simulations may raise ethical concerns regarding individual privacy and societal interventions. It is crucial to exercise caution in how the simulation results are handled.

# 6. Hypotheses for Social Simulations Inspired by Graphene's Properties

The unique properties of graphene, such as its robust structure and excellent electrical conductivity, can be applied as analogies in social phenomena, such as efficiency of information transfer and stability of social structures. For instance, by modeling graphene's hexagonal lattice structure as a social network and quantifying the strength of connections and efficiency of information flow, insights into optimal structures of social systems and optimization of information transmission may be obtained.

A hypothesis could be that "social networks resembling the lattice structure of graphene exhibit superior efficiency and resistance in information transmission". Through such models, principles for designing efficient and stable social systems could be explored.

To analyze the metaphorical relationship between the structural features of tellurium nanoparticles (TeNPs) and the societal group responses to fake news dissemination, we propose the following steps. This metaphorical analysis aims to provide insights for understanding the mechanisms of fake news spread and devising prevention strategies. Strengthening Covalent Bonds: The strengthening of covalent bonds in TeNPs corresponds to increased cohesion among individuals within a societal group. This cohesion can be modeled as the strength of shared beliefs and values.

Measure the similarity in beliefs and values among individuals within societal groups or subgroups and use it as an indicator of cohesion. Higher similarity corresponds to stronger covalent bonds, indicating a unified group. Disintegration of Secondary Structures: The process of disintegration of the secondary structures' inter-chain interactions in TeNPs corresponds to the disconnection of information flow between subgroups with different beliefs and values. Use network analysis to measure the degree of information exchange between subgroups and evaluate the extent of disconnection. Less flow of information implies a greater disconnection.

Introduction of Fake News: Introduce fake news into the societal group and observe how it resonates and spreads within the group. If fake news strongly resonates with a specific subgroup, that subgroup is particularly sensitive to fake news.

Measure the acceptance level of fake news within each subgroup and analyze which subgroups are more sensitive to fake news. Formation of Isolated Chains: Analyze the process by which subgroups susceptible to fake news become isolated and cling to fake news. This corresponds to the isolated chain-like structures in TeNPs.

Identify individuals or subgroups that continue to support fake news over a long period and measure their degree of stubbornness.

Simulation: Use the model constructed through the above steps to simulate the spread patterns of fake news within societal groups.

Extracting Insights: Analyze the data obtained from the simulation to extract insights on the mechanisms of fake news spread and prevention strategies.

This approach provides a metaphorical framework for understanding the spread and response of fake news within societal groups using the structural features of TeNPs. Yes, we propose the following specific equations and calculation processes:

1. Plasmon Resonance and Information Resonance

The strength of plasmon resonance is determined by the dielectric function of TeNPs and the dielectric constant of the surrounding medium. To represent information resonance, the following equation can be used:

``` $\epsilon(\omega) = \epsilon_\infty - \frac{\omega_p^2}{\omega^2 + i\gamma\omega}$ ```

Here, - $\epsilon(\omega)$ is the dielectric function at frequency $\omega$ - $\epsilon_\infty$ is the high-frequency limit dielectric constant - $\omega_p$ is the plasma frequency - $\gamma$ is the damping constant

The resonance strength $R_{ij}$ of fake news $j$ in subgroup $i$ can be defined as follows:

``` $R_{ij} = |\epsilon_i(\omega_j) - \epsilon_m|^{-2}$ ```

Here, - $\epsilon_i(\omega_j)$ is the dielectric function of subgroup $i$ evaluated at the characteristic frequency $\omega_j$ of fake news $j$ - $\epsilon_m$ is the dielectric constant of the surrounding medium (the whole society)

A higher $R_{ij}$ indicates that subgroup $i$ strongly resonates with fake news $j$.

2. Snell's Law and Information Refraction Reflection

The refraction and reflection of information between different subgroups can be considered to follow Snell's Law:

''' $n_1 \sin(\theta_1) = n_2 \sin(\theta_2)$ '''

Here, - $n_1, n_2$ are the "refractive indices" of subgroups 1 and 2, respectively - $\theta_1, \theta_2$ are the angles of incidence and refraction, respectively

The "refractive index" $n_i$ of subgroup $i$ can be set according to the strength of beliefs and values within that subgroup.

The reflection rate $R$ can be calculated using the following equation:

''' $R = \left(\frac{n_2 - n_1}{n_2 + n_1}\right)^2$ '''

A higher reflection rate indicates that information is less likely to be transmitted between subgroups.

3. Surface Plasmon Resonance and Information Propagation

Information propagation across the boundary between subgroups can be derived from the model of surface plasmon resonance. The propagation constant $\beta$ is given by:

''' $\beta = \frac{\omega}{c} \sqrt{\frac{\epsilon_1 \epsilon_2}{\epsilon_1 + \epsilon_2}}$ '''

Here, - $\omega$ is the characteristic frequency of information - $c$ is the speed of light - $\epsilon_1, \epsilon_2$ are the dielectric functions of subgroups 1 and 2, respectively

The propagation distance $L$ can be calculated as:

''' $L = 1 \frac{}{\text{Im}(\beta)}$ '''

A larger $L$ indicates that information can easily propagate across the boundary between subgroups.

# 7. Proposal for Analyzing Fake News Diffusion

Based on an approach that analyzes the diffusion of fake news and social isolation similar to the properties of tellurium nanoparticles and graphene, we propose specific equations and computational processes as follows:

## 7.1 Modeling Information Resonance

To model the resonance of information (including fake news) within a social group, we apply the equation for plasmon resonance. Each member of the social group (individual) is considered as a nanoparticle, and the resonance of information between them is considered.

(1) We denote the information receptivity of individual $i$ as $R_i(\omega)$, which can be expressed by the following equation:

$$R_i(\omega) = |\epsilon_i(\omega) - \epsilon_m|^{-2}$$

where:

$\epsilon_i(\omega)$ represents the "dielectric function" of individual $i$ at the information resonance frequency,

$\epsilon_m$ denotes the average "dielectric function" of the media or the entire society.

(2) Using the information receptivity of each individual, we quantify the resonance strength of information at the group level as follows:

$$R_{\text{group}}(\omega) = \frac{1}{N} \sum_{i=1}^{N} R_i(\omega)$$

where $N$ is the number of individuals in the group.

## 7.2 Refraction and Reflection of Information

The transmission of information between social groups is modeled using Snell's law. We consider the refraction (transmission) and reflection (rejection or misunderstanding) of information between different social groups.

(1) Let $n_A$ and $n_B$ represent the refractive indices of social groups A and B, respectively. The reflection coefficient $R_{AB}$ when information is transmitted from group A to group B is given by:

$$R_{AB} = \left(\frac{n_B - n_A}{n_B + n_A}\right)^2$$

where the refractive indices are defined based on the information receptivity and openness of the groups.

(2) Using the reflection coefficient, we define the transmission efficiency $T_{AB}$ of information as follows:

$$T_{AB} = 1 - R_{AB}$$

## 7.3 Analysis of Social Isolation

To model social isolation caused by the diffusion of fake news, we calculate the degree of isolation based on information resonance and transmission efficiency.

(1) Let $I_A$ denote the degree of isolation of group A, which is defined based on the resonance strength to fake news and the low efficiency of information transmission from other groups:

$$I_A = R_{\text{group},A}(\omega) \times \left(1 - \sum_{\text{all } B \neq A} T_{AB}\right)$$

where $R_{\text{group},A}(\omega)$ is the resonance strength of group A to fake news, and $T_{AB}$ is the efficiency of information transmission between group A and other groups.

(2) If the degree of isolation within a group exceeds a certain threshold, the group is more susceptible to the influence of fake news and tends to be isolated from the society as a whole.

Through the application of models based on the plasmonic properties of tellurium nanoparticles and graphene, it is possible to analyze the social impact of fake news and the dynamics of information transmission between groups. This is expected to contribute to understanding the diffusion mechanism of fake news and formulating strategies to address it.

# 8. Proposal for Analyzing the Relationship Between Structural Features of Tellurium Nanoparticles and the Spread of Fake News

To analyze the metaphorical relationship between the structural features of tellurium nanoparticles and the societal responses to the spread of fake news, we propose the following steps. This metaphorical analysis aims to provide insights into understanding the mechanisms of fake news diffusion and devising preventive measures.

## 8.1 Strengthening of Shared Bonds and Social Cohesion

(1) Strengthening of Shared Bonds: The enhancement of shared bonds in tellurium nanoparticles corresponds to the strengthening of interpersonal cohesion within a social group. This cohesion can be modeled as the strength of shared beliefs or values.

(2) Quantification of Cohesion: Measure the similarity of beliefs or values among individuals within each social group or subgroup and use it as an indicator of cohesion. High similarity corresponds to strong shared bonds, indicating a sense of unity within the group.

## 8.2 Collapse of Secondary Structure and Information Disruption

(1) Collapse of Secondary Structure: The collapse of inter-chain interactions in tellurium nanoparticles corresponds to the disruption of information flow between subgroups with different beliefs or values.

(2) Quantification of Information Disruption: Measure the degree of information exchange between subgroups through network analysis and evaluate the extent of disruption. Less flow of information implies greater disruption.

## 8.3 Impact and Resonance of Fake News

(1) Introduction of Fake News: Introduce fake news into social groups and observe how it resonates and spreads within the groups. If fake news resonates strongly with a particular subgroup, that subgroup can be considered particularly sensitive to fake news.

(2) Quantification of Resonance: Measure the receptivity of fake news within each subgroup and analyze which subgroups are most sensitively responsive to fake news.

## 8.4 Analysis of Isolated Chain-like Persistence

(1) Formation of Isolated Chains: Analyze the process by which subgroups susceptible to the influence of fake news become isolated and persist in adhering to fake news. This corresponds to the chain-like isolated structure of tellurium nanoparticles.

(2) Quantification of Persistence: Identify individuals or subgroups that continue to support fake news over an extended period and measure the degree of their persistence.

## 8.5 Validation of the Model and Insights

(1) Simulation: Simulate the diffusion patterns of fake news within social groups using the model constructed through the above steps.

(2) Extraction of Insights: Analyze the data obtained from simulations and extract insights into the mechanisms of fake news diffusion and preventive measures.

This approach provides a metaphorical framework for understanding the spread of fake news and responses within social groups, using the structural features of tellurium nanoparticles.

# 9. Modeling the Enhancement of Shared Bonds

To model the cohesion within a social group as the strengthening of shared bonds, the following steps are taken.

## 1. Definition of Interpersonal Relationships within the Group

Each individual within the group is represented as a node, and relationships between individuals (e.g., friendships, colleague relationships) are represented as edges in a graph.

The weight $w_{ij}$ of an edge between individuals $i$ and $j$ represents the strength of their relationship or the similarity of their beliefs. The weight is quantified based on the intensity of shared beliefs or values.

## 2. Calculation of Shared Cohesion $B_s$

The cohesion within the group $B_s$ is calculated as the average weight of all edges within the group.

$$B_s = \frac{1}{N} \sum_{i=1}^{N} \sum_{j=1}^{N} w_{ij}$$

Here, $N$ is the total number of individuals in the group, and $w_{ij}$ is the weight of the edge between individuals $i$ and $j$.

# Modeling the Variation of Shared Cohesion

## 1. Changes in Cohesion Due to External Influences

The cohesion within the group may change over time due to the introduction of new information or fake news. This change can be modeled by introducing the time dependence of $B_s(t)$.

The change in cohesion $\Delta B_s$ at time $t$ can take positive or negative values depending on the strength and direction of the external influences.

## 2. Calculation of Temporal Changes in Cohesion

The update of cohesion at time $t$ is calculated using the following equation.

$$B_s(t+1) = B_s(t) + \Delta B_s$$

Here, $\Delta B_s$ is the change in cohesion due to external influences at time $t$.

# Modeling the Collapse of Secondary Structure

To model the collapse of secondary structure (interactions between subgroups) within a social group, the following steps are taken.

## Identification of Subgroups

Subgroups within the social group are identified. This can be done using community detection algorithms.

Each subgroup is defined as a gathering of individuals who share similar beliefs or interests.

## Quantification of Interactions between Subgroups

The strength of interactions between subgroups is quantified using edge weights $w_{kl}$. Here, $k$ and $l$ denote individuals belonging to different subgroups.

The strength of interactions is quantified based on the amount and quality of information shared between subgroups.

## Quantification of Collapse of Secondary Structure

The collapse of secondary structure is considered to occur when the strength of interactions between subgroups falls below a certain threshold.

This collapse can be quantified by calculating the "disruption degree" $D(t)$ using the following formula.

$$D(t) = \frac{1}{M} \sum_{k=1}^{M} \sum_{\substack{l=1 \\ l \neq k}}^{M} (1 - w_{kl})$$

Here, $M$ is the total number of subgroups, and $w_{kl}$ is the strength of interaction between subgroups $k$ and $l$. A higher value of $D(t)$ indicates weaker interactions.

## 1. Construction of Social Network

Build the network of the social group and identify subgroups. The weight of each edge represents the strength of interaction between subgroups.

## 2. Calculation of Disruption Degree

Calculate the strength of interactions between each pair of subgroups at time $t$, and obtain the disruption degree $D(t)$ by taking the average of all pairwise interactions.

## 3. Analysis of Collapse of Secondary Structure

Track the disruption degree $D(t)$ over time and analyze the progress of the collapse of secondary structure. Observe changes in the disruption degree due to the introduction of fake news or external influences.

# 10. Modeling the Impact and Resonance of Fake News

The proposed equations and calculation process for modeling the impact of fake news and resonance are as follows:

## Modeling the Impact of Fake News

To model the impact of fake news on individuals or social groups, it is necessary to quantify how fake news "resonates" with individuals' beliefs and attitudes.

(1) **Definition of Resonance**:

Resonance is quantified as the degree to which an individual's current beliefs or values align with the content of fake news. It serves as an indicator of the receptivity and potential spread of fake news.

(2) **Setting up the Resonance Function**:

Let $R_i$ denote the resonance of individual $i$ with fake news, defined as follows:

$$R_i = \exp\left(-\frac{(B_i - F)^2}{2\sigma^2}\right)$$

where $B_i$ represents the strength of individual $i$'s beliefs, $F$ represents the content of fake news, and $\sigma$ is a parameter controlling the spread of resonance.

(1) **Initialization of Individual Beliefs**:

Initialize the beliefs $B_i$ for each individual in the social group. This can be done randomly or based on actual data.

(2) **Introduction of Fake News**:

Introduce fake news $F$ at a certain point in time and compute the resonance $R_i$ for all individuals in the social group.

(3) **Updating Beliefs**:

Update individuals' beliefs based on the resonance of fake news. The belief update is performed as follows:

$$B'_i = B_i + \alpha R_i$$

where $\alpha$ is a parameter controlling the strength of the update.

(4) **Analysis of Dynamic Changes in Resonance and Beliefs**:

Track changes in individuals' beliefs over time and analyze the impact of fake news resonance on beliefs.

This calculation process involves identifying subgroups using community detection algorithms, computing the disruption degree $D(t)$, and visualizing the results. This allows us to understand how fake news affects the secondary structure of a social group.

Fig.3 provided is a color-coded representation of the belief state of a node at time step 10 in a given network. The colors here indicate the belief state of a node, with red meaning a high belief state (high probability of believing fake news) and blue meaning a low belief state (low probability of believing fake news).

As a consideration from the isolation of the telluric chain, it can be seen that some nodes in the network have lower belief states than others in this graph. These blue nodes may suggest that the fake news information is not propagated very well or is completely isolated. If these nodes have sources or beliefs that are resistant to fake news, their isolation may have a protective effect against the effects of fake news.

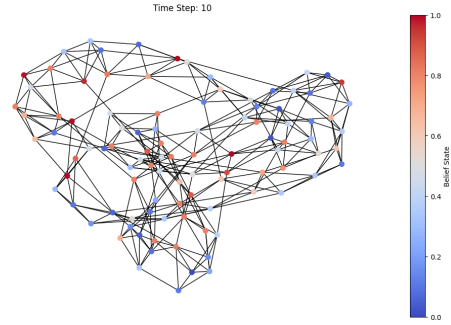

Fig. 3: Network Belief States susceptibility

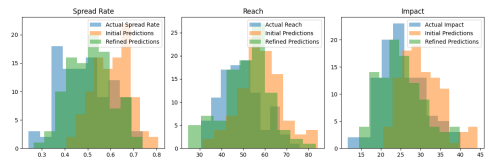

Fig. 4: Actual Reach, Speed, Spread

On the other hand, in terms of isolation of excessive fake news spreaders, it is important to look at whether red nodes have many connections. Red nodes that are centrally located have many connections, thereby indicating that they are more likely to spread fake news. However, if these nodes are on the periphery of the network or have few connections, their spreading power is limited. In the graph, some red nodes appear to have only a few connections. If these nodes have a strong belief in fake news but are isolated, this suggests that their impact may not extend to the entire network.

Fig.4 provided are histograms showing actual data on the Spread Rate, Reach, and Impact of fake news, as well as initial and improved predictions for them.

In terms of telluric chain isolation, if the actual measured values of the spread rate, reach, and impact of fake news are lower than the initial and improved predictions, this phenomenon may indicate that some agents in the network are cut off from the information flow, resulting in less spread of fake news than expected There is a possibility that this may indicate that some agents in the network are cut off from the flow of information. This may mean that certain populations are more resistant to fake news or are protected by accurate information from reliable sources.

In terms of isolation of excessive fake news spreaders, if the measured value is higher than predicted, it is possible that influential individuals in the network are actively spreading fake news. However, if this diffusion is more limited than expected, it may indicate that the spreaders of fake news are isolated within the network or that their message is not as well accepted as expected.

Overall, the lower than expected actual diffusion rate,

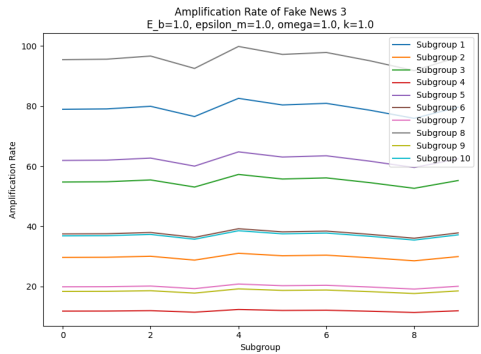

Fig. 5: Network Belief States after Fake News Propagation

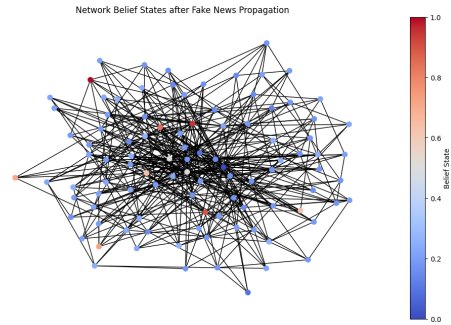

Fig. 6: Amplification Rate of Fake News

reach, and impact suggests that a telluric chain effect exists in the social network, creating a population that is more resistant to fake news. This could be the result of improved information literacy, better access to verifiable sources, or effective communication strategies by the community.

On the other hand, where the measured values exceed the predicted values, it means that the spreaders of fake news have some influence, and it is important to use this information to conduct network analysis to further investigate how these spreaders are isolated or connected. This will help to more accurately understand the flow of information and take steps to combat fake news.

Fig.5 provided shows the amplification rate of fake news in different subgroups. Each subgroup is represented by a different colored line, with the vertical axis representing the amplification rate and the horizontal axis representing the subgroup.

Taking into account the isolation of the telluric chain, some subgroups (e.g., subgroup 10) have low amplification rates, which may suggest that they are relatively protected from the effects of fake news or that they are resistant to the acceptance of fake news. On the other hand, subgroups with high amplification (e.g., subgroup 1) imply that fake news is spread more effectively.

In terms of isolation of excessive fake news spreaders, a subgroup with a high amplification rate may indicate that within that subgroup the fake news spreader has a greater influence on other members. This may indicate that the spreader plays a central role, or it may indicate a low tolerance for fake news within the subgroup.

If there is a large difference in amplification between subgroups, it means that different subgroups are responding differently to fake news. This could vary depending on the strength of communication within the subgroup, access to reliable sources, or awareness of fake news.

This information can be useful in understanding how susceptible a particular subgroup is to fake news. It also provides insight to determine which subgroups to focus on and what approach to take when planning strategies to prevent the spread of fake news. Subgroups with particularly high amplification rates need to understand why and take specific measures to address them. At the same time, learning from subgroups with low amplification will help you find ways to increase your resistance to fake news.

Fig.6, Shows the belief state of the network after the propagation of fake news. The color shade represents the belief state of each node, with blue meaning a low belief state and red meaning a high belief state. This provides a visual understanding of how fake news is accepted and propagated in the network.

With respect to the isolation of telluric chains, we can assume that nodes in the network with low belief states (blue nodes) are less susceptible to fake news. These nodes may have few or no connections to agents spreading fake news. A telluric chain may refer to the phenomenon of a particular population being disconnected or isolated from the social network. In this case, isolated nodes may be protected from information propagation and at a lower risk of being affected by fake news.

On the other hand, in terms of isolation of excessive fake news spreaders, it is suggested that nodes with strong fake news propagation power (red nodes) spread fake news to other nodes through their influence. If these nodes are spreading fake news excessively, their position in the network is important. Red nodes that are centrally located can influence more nodes due to their high connectivity. On the other hand, if red nodes are located on the periphery of the network, their influence will be limited and they may be considered isolated.

Based on this code and graph, the nodes most susceptible to fake news are likely to be located near the source of the fake news. Nodes with high susceptibility are also expected to be more susceptible to fake news. Nodes whose belief state has been raised to 1.0 (fully believe) by fake news will influence their surrounding nodes to spread the belief in fake news.

These observations help us understand how fake news spreads in the network and affects the belief state of individual nodes.

# 11. Perspect

When applying simulations based on first-principles calculations in the context of fake news propagation, several effective materials and phenomena other than TeNPs and graphene can be considered. Here are some examples:

- **Topological Insulators**: Topological insulators are special materials that, despite being insulating inside, exhibit conductivity on their surfaces. This surface state is topologically protected and highly stable against external disturbances. In social simulations, this property of topological insulators could be useful in mimicking situations where the internal structure of social groups is protected while vigorous information exchange with the external environment occurs. Particularly, it could be applied to represent the dynamics of societies where different opinions and values are retained internally while flexible dialogues are possible externally.

- **Two-Dimensional Materials**: Apart from graphene, there are many two-dimensional materials, each possessing unique electronic properties. Examples include transition metal dichalcogenides (TMDs), black phosphorus (phosphorene), hexagonal boron nitride (h-BN), etc. These materials exhibit distinct properties such as different band gaps and carrier mobilities, which can be used as parameters to mimic various societal processes related to information diffusion and the formation of filter bubbles.

- **Metamaterials and Photonic Crystals**: Metamaterials and photonic crystals are artificial materials exhibiting peculiar optical properties not found in nature. These materials demonstrate abnormal refractive indices and transmittance for specific wavelengths and directions of light. In social simulations, such properties of metamaterials could be applied to control mechanisms for selective transmission, reflection of information, i.e., formation and disruption of filter bubbles.

By utilizing these properties, it would be possible to provide new perspectives for understanding the complex interactions of social phenomena such as fake news propagation, dynamics of information exchange among social groups, and formation and dissolution of filter bubbles in greater detail.

## 11.1 First-principles calculations in physics:Specific Formulas and Computational Procedures

First-principles calculations in physics, especially in solid-state physics and materials science, refer to the method of directly calculating the properties of materials from the fundamental physical laws (mainly quantum mechanics). When considering their application to social simulations, one can explore using ideas and concepts obtained from first-principles calculations in materials science to analyze non-physical phenomena such as the diffusion of fake news among social groups.

## Analysis of Isolated Chain-Like Adherence

In the analysis of isolated chain-like adherence, analogous to the isolated chain-like properties of tellurium nanoparticles, the behavior of individuals or subgroups strongly adhering to fake news is examined. This step models how individuals with high resonance to fake news isolate information and maintain or strengthen their beliefs, akin to the isolated chain-like behavior of tellurium nanoparticles.

### Modeling Isolated Chain-Like Adherence

(1) **Definition of Adherence**:

The degree to which an individual adheres to fake news is defined as their "adherence" $P_i$. Adherence is calculated based on the strength of the individual's beliefs and the resonance with fake news.

(2) **Formula for Adherence**:

Adherence $P_i$ is calculated using the resonance with fake news $R_i$ and the individual's isolation degree $I_i$ (indicating reduced information exchange with others).

$$P_i = R_i \times I_i$$

where the isolation degree $I_i$ increases with a decrease in the number or frequency of information exchanges with others.

(1) **Measurement of Isolation Degree**:

For each individual, measure the number of others with whom they exchanged information within a certain period to determine the isolation degree $I_i$. This is calculated based on the frequency and extent of social interactions.

(2) **Calculation of Adherence**:

Calculate the adherence $P_i$ of each individual based on their isolation degree $I_i$ and resonance with fake news $R_i$.

(3) **Analysis of Adherence**:

Identify individuals or subgroups within the social group with high adherence and analyze their attitudes and behaviors towards fake news.

(4) **Tracking Changes Over Time**:

Track changes in individual adherence over time and observe the development of isolated chain-like adherence to fake news.

0 **Electronic Structure Calculation**:

In first-principles calculations of materials properties, it often starts with solving the Schrödinger equation or its approximation, such as Koopmans' theorem. The Schrödinger equation is given as:

$$\hat{H}\Psi = E\Psi$$

where $\hat{H}$ is the Hamiltonian (an operator representing the total energy of the system), $\Psi$ is the wave function, and $E$ is the energy eigenvalue. Solving this equation allows obtaining the electronic structure of materials.

Application: This analogy can be used to analyze the transition probabilities and stability of social groups or subgroups, considering them to be in different "energy states", to understand decision-making processes within social groups or the acceptance of fake news.

0 **Density Functional Theory (DFT)**:

DFT is a commonly used method in first-principles calculations, treating the electron density as the fundamental variable. The basic idea is to construct the total energy of the system, represented by the total electron density $\rho(\mathbf{r})$, as an energy functional $E[\rho]$, and minimize it to determine the properties of the ground state.

Application: By considering the "density" of opinions or information among subgroups and its influence on the overall "energy state", one can analyze the efficiency of information flow within social groups or the tendency of fake news diffusion.

0 **Path Integral Monte Carlo Method**:

The path integral Monte Carlo method is a technique for computing properties of quantum many-body systems, where the partition function of the quantum system is obtained through path integration, followed by statistical sampling. This method is particularly useful for systems where quantum effects are significant at low temperatures.

Application: Modeling decision-making processes within social groups or the diffusion patterns of fake news as "quantum-like fractal paths" and using the probabilistic nature of these paths to capture the complexity of social phenomena.

When applying these physics methods to social science problems, it is essential to appropriately map physical concepts to the social science context. For example, "energy" can represent the level of activity or influence within a social group, and "electron density" can serve as a metaphor for the density of opinions or information distribution. By utilizing such analogies, it may lead to a new understanding of social phenomena and the construction of predictive models.

# Analysis of Isolated Chain-Like Adherence

In the analysis of isolated chain-like adherence, analogous to the isolated chain-like properties of tellurium nanoparticles, the behavior of individuals or subgroups strongly adhering to fake news is examined. This step models how individuals with high resonance to fake news isolate information and maintain or strengthen their beliefs, similar to the isolated chain-like behavior of tellurium nanoparticles.

## Modeling Isolated Chain-Like Adherence

**Introduction of Plasmonic Resonance**

By introducing plasmonic resonance and Snell's law into this scenario, new insights into the diffusion of fake news and the reaction of social groups can be provided. Plasmonic resonance is the phenomenon of light resonating on nanoparticles or metal surfaces, and Snell's law describes the refraction of light. Applying these physical concepts to the context of social science can deepen understanding of the diffusion of fake news and interaction within social groups.

(1) **Introduction of Plasmonic Resonance**:

Resonance and sharing of beliefs: Plasmonic resonance in tellurium nanoparticles corresponds to the process where fake news resonates with specific beliefs or values within a social group. When fake news strongly resonates with the beliefs of a particular group, its dissemination within that group is facilitated.

Strength of resonance: The strength of resonance is modeled as the degree to which fake news is accepted by the social group. Strong resonance implies high credibility or persuasiveness of fake news, or alignment with existing beliefs.

(2) **Application of Snell's Law**:

Refraction of information: Snell's law is metaphorically applied to describe the "refraction" of information between different subgroups. When information is transmitted between groups with different belief systems, it may be distorted or misunderstood.

Angle of information: The "angle" of information transmission (analogous to the angle of incidence in Snell's law) indicates how well the information aligns with the beliefs of the receiving group. A smaller angle (i.e., information transmitted aligns with the beliefs of the receiving group) results in less distortion in acceptance of the information.

**Analytical Ideas**

(1) **Construction of Resonance Model**:

Quantify the belief systems of each social group or subgroup and construct a model to measure resonance with fake news. This model can help predict how effectively fake news spreads within specific groups.

(2) **Simulation of Information Transmission**:

Simulate information transmission between groups with different belief systems and analyze how distortion or misunderstanding of information occurs. This process helps understand how fake news is interpreted and reconstructed within the social group.

(3) **Proposal of Preventive Measures**:

Based on the resonance model and simulations of information transmission, propose strategies to prevent the spread of fake news. For example, communication methods to minimize information "refraction" or educational programs to weaken resonance within the social group can be considered.

## 11.2 Discussion of the structural characteristics of tellurium nanoparticles and their application to social sciences

In the discussion of the structural characteristics of tellurium nanoparticles and their application to social sciences, it is essential to consider both the challenges and benefits when applying research findings from the physical sciences to the analysis of social phenomena.

0 **Limits of Analogies**:

Direct comparison between the structural properties of physical nanoparticles and the dynamics of social groups has limitations due to inherent differences. While the collapse of tellurium nanoparticle bonds and interchain interactions is a direct physical process, the sharing of beliefs and the discontinuity of information within social groups are shaped by more complex psychological and social factors.

0 **Complexity of Models**:

Models in the physical sciences are often mathematically rigorous, allowing for quantitative predictions. However, phenomena in social sciences are strongly influenced by the uncertainty of human behavior and decision-making, making it challenging to directly apply these models.

0 **Availability of Data**:

While precise data can be obtained using advanced experimental facilities in the study of tellurium nanoparticles, research on social groups may be limited in data collection due to privacy concerns and ethical constraints.

0 **Offering New Perspectives**:

Analogies from the physical sciences provide new perspectives on issues in social sciences. By applying the structural characteristics of tellurium nanoparticles to social groups, it becomes possible to understand phenomena such as the diffusion of fake news and information discontinuity from a fresh angle.

0 **Enhancing Modeling and Prediction**:

By referencing the precise mathematical models from physics, the accuracy of modeling in social sciences research can be improved. This may be particularly useful for quantitatively analyzing the mechanisms of fake news diffusion and predicting the behavior of social groups.

0 **Designing Strategic Interventions**:

Using models inspired by the structure of tellurium nanoparticles, more effective strategies to counteract fake news can be designed. For example, interventions aimed at strengthening cohesion within social groups or resolving information discontinuity can be considered.

In conclusion, while applying the structural characteristics of tellurium nanoparticles to social sciences presents certain challenges, it holds the potential to provide valuable insights into understanding the mechanisms of fake news diffusion and devising preventive measures. When using analogies, it is important to fully understand the limitations of their application and interpret them carefully within the context of social sciences.